
\input phyzzx
\tolerance=10000

{\hsize=17.5truecm \leftskip=9.5cm
{NDA-FP-4/92, OCHA--PP-26}\par
\vskip -4mm
{June 1992}\par}

\def\cref#1{\rlap,\attach{#1}}
\def\pref#1{\rlap.\attach{#1}}

\def\ie{{\it i.e.}}

\def\e{{\rm e}}

\def\sinh{{\rm sinh\,}}
\def\tanh{{\rm tanh\,}}
\def\half{{1\over 2}}

\def\pint{{1 \over 2\pi}\int d^2x\,}
\def\sqg{\sqrt{-g}}
\def\ephi{\e^{-2\phi}}
\def\erho{\e^{-2\rho}}
\def\epphi{\e^{2\phi}}
\def\eprho{\e^{2\rho}}
\def\ephirho{\e^{2(\phi-\rho)}}
\def\RN{Reissner-Nordstr\"om}

\title{Charged Dilatonic Black Hole and Hawking Radiation in Two Dimensions}
\author{Shin'ichi Nojiri}
\address{Department of Mathematics and Physics}
\address{National Defense Academy}
\address{Hashirimizu, Yokosuka 239, JAPAN}

\author{Ichiro Oda}
\address{Faculty of Science, Department of Physics}
\address{Ochanomizu University}
\address{1-1 Otsuka 2, Bunkyo-ku, Tokyo 112, JAPAN}

\abstract{We consider Callan, Giddings, Harvey and Strominger's (CGHS)
two dimensional dilatonic gravity with electromagnetic interactions.
This model can be also solved classically.
Among the solutions describing static black holes,
there exist extremal solutions which have zero temperatures.
In the extremal solutions, the space-time metric is not singular.
We also obtain the solutions describing charged matter (chiral fermions)
collapsing into black holes.
Through the collapsing, not only future horizon but past horizon
is also shifted.
The quantum corrections including chiral anomaly are also discussed.
In a way similar to CGHS model, the curvature singularity also appeared,
except extremal case, when the matter collapsing.
The screening effects due to the chiral anomaly have a tendency to
cloak the singularity.}

\endpage
\REF\i{S.W. Hawking\journal Comm.Math. Phys. &43 (75) 199}
\REF\v{G. 't Hooft\journal Nucl.Phys. &B335 (90) 138}
\REF\ii{Y. Aharonov, A. Casher and S. Nissinov\journal Phys.Lett. &B191
(87) 51}
\REF\iv{C.F.E. Holzhey and F. Wilczek, preprint IASSNS-HEP-91/71}
\REF\iii{C.G. Callan, S.B. Giddings, J.A. Harvey and A. Strominger\journal
Phys.Rev. &D45 (92) R1005}
\REF\vi{J.G. Russo, L. Susskind and L. Thorlacius,
preprint SU-ITP-92-4 (1992)}
\REF\vii{L. Susskind and L. Thorlacius,
preprint SU-ITP-92-12 (1992)}
\REF\xi{S.W. Hawking, preprint CALT-68-1774}
\REF\xii{T. Banks, A. Dabholkar, M.R. Douglas
and M. O'loughlin\journal Phys.Rev. &D45 (92) 3607}
\REF\viii{N. Ishibashi, M. Li
and R. Steif\journal Phys.Rev.Lett. &67 (91) 3336}
\REF\ix{M.D. McGuigan, C.R. Nappi
and S.A. Yost\journal Nucl.Phys. &B375 (92) 421}
\REF\x{S. Coleman, J. Preskill and F. Wilczek, preprint LASSNS-HEP-91/17
(1991)}
\REF\xiii{S. Nojiri and I. Oda, in preparation}

When we try to construct the quantum gravity theory, black hole
evaporation provides a serious problem. A quantum mechanically pure state
which describes gravitationally collapsing matter to form a black
hole, evolves into a mixed quantum state, which describes Hawking
radiation\pref\i
There are several scenarios to solve this problem of the loss of quantum
informations.
Hawking gave a most radical proposal that the black hole completely
evaporates and the quantum coherence is lost in the gravitational
collapse.
On the other hand, 't Hooft has proposed that Hawking radiation
carries off informations about the quantum states of the black
holes\pref\v
Another conservative proposal is that the process of the collapse and
the radiation leaves a stable remnant which carries the information of
the initial configuration of the system\pref\ii
However, if the remnant has the mass of order the Planck scale,
gravitational effects would produce the light remnants in pair and
the lifetime of stars would be shorter than observed.
Therefore the mass of the remnants should be macroscopic.
A candidate of such remnants is the extremal black holes of
\RN\ solutions\pref\iv The extremal black holes have
vanishing temperatures and the space-time metric is not singular everywhere.

Recently, Callan, Giddings, Harvey and Strominger have investigated
an interesting toy
model of two dimensional gravity\pref\iii
In this model, the gravity couples with a dilaton and conformal matter
fields. They found classical exact solutions, including the solutions
describing the formation of a black hole by collapsing matter.
Furthermore Hawking radiation and the back reaction of the metric can be
described by adding correction terms to the classical equations of motion.
The correction terms derive from the conformal anomaly of the matter
field.
It was also argued that the quantum correction would remove the usual
black hole singularity.
Several groups\cref{\vi,\xi,\xii} however, pointed that gravitational
collapse always leads to a curvature singularity.
Susskind and Thorlacius have found massive static solutions with zero
temperature but they have shown that a decaying black hole cannot evolve into
one of them.

In this paper, we couple electromagnetic fields to CGHS
dilatonic gravity in a special way.
This model can be also solved classically.
Among the solutions describing static black holes,
there exist extremal solutions which have zero temperatures.
In the extremal solutions, the space-time metric is not singular
and these solutions should be analogues of four dimensional extremal
\RN\ black holes.
We also obtain the solutions describing charged matter (chiral fermions)
collapsing into black holes.
Through the collapse, not only future horizon but past horizon
is also shifted.
We can also discuss the quantum corrections by including chiral anomaly.
When the matter collapses into a black hole,
except the case that the resulting black hole is extremal,
the curvature singularity also appeared in the quantum theory
in a way similar to CGHS model.

We start with the following action,
$$S_c=\pint\sqg\{\ephi(R+4(\nabla \phi)^2+4\lambda^2)
-{\e^{a\phi} \over g_A^2}F^2
-\sum_{j=1}^N i\bar\Psi_j\gamma^\mu(D_\mu-iA_\mu)\Psi_j\}
\ . \eqn\ei$$
Here $\phi$ is a dilaton field and $\Psi_j={\psi_j \choose 0}$'s are
$N$ left-handed complex fermions. $g_A$ is a $U(1)$ electromagnetic gauge
coupling constant and $D_\mu$ is a covariant derivative.
If we fix $a=-2$, this action, except the fermion part, describes
an effective field theory derived from string theory\pref{\viii, \ix}
In this paper, we choose $a=2$, which allows us to exactly solve this model
classically.
When we fix the gauge freedom of reparametrization invariance
by the conformal gauge,
$$g_{\mp\pm}=-\half\eprho\ , \ \ g_{\pm\pm}=0\ , \ \eqn\eii$$
and that of $U(1)$ gauge symmetry by the following light-cone gauge condition
$$A_-=0 \ , \eqn\eiib$$
the action \ei\ reduces into
$$\eqalign{S_c&=\pint\{\ephi(4\partial_+\partial_-\rho
-8\partial_+\phi\partial_-\phi+2\lambda^2\eprho)\cr
&\ \ \ \ +{4 \over g_A^2}\e^{2\phi-2\rho}F_{+-}^2
+{i \over 2}\sum_{j=1}^N\psi_j^*\partial_-\psi_j\}
\ .} \eqn\eiii$$
Here $F_{+-}=-\partial_-A_+$.
The metric equations are given by
$$\eqalign{0&=T_{++}=\ephi(4\partial_+\rho\partial_+\phi-2\partial_+^2\phi)
+{i \over 4}\sum_{j=1}^N(\psi_j^*\partial_+\psi_j-\partial_+\psi_j^*\psi_j)\cr
&\ \ \ \ \ \ \ +\half A_+\sum_{j=1}^N\psi_j^*\psi_j \ ,
\cr
0&=T_{--}=\ephi(4\partial_-\rho\partial_-\phi-2\partial_-^2\phi)\ ,}
\eqn\ev$$
$$0=T_{+-}=\ephi(2\partial_+\partial_-\phi
-4\partial_+\phi\partial_-\phi-\lambda^2\eprho)
+{2 \over g_A^2}\e^{2\phi-2\rho}F_{+-}^2
\ . \eqn\evi$$
The Maxwell equations are
$$\eqalign{0&=-{8 \over g_A^2}\partial_+(\ephirho F_{+-})
+\half\sum_{j=1}^N\psi_j^*\psi_j\ ,\cr
0&=-{8 \over g_A^2}\partial_-(\ephirho F_{+-})\ .}
\eqn\evii$$
We also obtain dilaton and matter equations of motion,
$$0=-4\partial_+\partial_-\phi+4\partial_+\phi\partial_-\phi
+2\partial_+\partial_-\rho+\lambda^2\eprho
-{2 \over g_A^2}\e^{4\phi-2\rho} F_{+-}^2
\ , \eqn\eviii$$
$$0=\partial_-\psi_j \ .\eqn\eix$$

The general solutions of Equations \ev\ - \eix\ are given by,
$$\eqalign{\psi_j&=\psi_j(x^+)\ ,\cr
A_+&=-{g_A^2 \over 4}x^-\int^{x^+}dy^+\sum_{j=1}^N\psi_j^*(y^+)\psi_j(y^+)
\ ,\cr
\ephi&=\erho={M \over \lambda}-\lambda^2x^+x^-
+{g_A^2 \over 8}x^-
\int^{x^+}dy^+(\int^{y^+}dz^+\sum_{j=1}^N\psi_j^*(z^+)\psi_j(z^+))^2\cr
&\ \ \ \
-{i \over 4}\int^{x^+}dy^+
\int^{y^+}dz^+\sum_{j=1}^N(\psi_j^*(z^+)\partial_+\psi_j(z^+)
-\partial_+\psi_j^*(z^+)\psi_j(z^+))\ .}
\eqn\ex$$
Here we have fixed the residual gauge symmetry of reparametrization by
choosing the condition $\phi-\rho=0$, which is given by a sum of
holomorphic function $w^+(x^+)$ and anti-holomorphic function $w^-(x^-)$
in general :
$\phi(x^+,x^-)-\rho(x^+,x^-)=w^+(x^+)+w^-(x^-)$.

Now we consider solutions with $\psi_j=0$.
A special solution describing a dilaton vacuum is given by,
$$A_+=0, \ \ \ephi=\erho=-\lambda^2x^+x^-\ .
\eqn\exa$$
We have more general solutions,
$$\eqalign{F_{+-}&=C \ \ ({\rm constant})\ .\cr
\ephi&=\erho={M \over \lambda}-\tilde\lambda^2x^+x^-\ .}
\eqn\exi$$
Here ${M \over \lambda}$ is an integration constant
and $\tilde\lambda$ is defined by
$$\tilde\lambda^2\equiv \lambda^2-{2C^2 \over g_A^2}\ . \eqn\exiii$$
The solutions \exi\ tell that the metric has a singularity when
${M \over \lambda}-\tilde\lambda^2x^+x^-=0$ and there
appear horizons $x^+x^-=0$. Therefore, if $C\neq 0$, the solutions \exi\
describe charged black holes. The structure of space-time is similar
to that of the Schwarzschild black holes and simple in contrast to
the \RN\ black hole solutions in four dimensions.
Note that the singularity vanishes when $\tilde\lambda^2=0$
when we fix $M$ to be finite,
although the solution corresponds to a massive\foot{
Here ``massive" means that ${M \over \lambda}$ does not vanish.}
and charged object.
This solution could be a natural anologue of extremal
\RN\ black hole solution.
In fact this solution has a vanishing temperature and this solution could
be a candidate of solutions describing black hole remnants\pref\vi
The temperature can be found by changing the coordinates
$$\eqalign{x^+&={1 \over \tilde\lambda}\sqrt{{M \over \lambda}}
\e^{\tilde\lambda t }\sinh \tilde\lambda r\cr
x^-&=-{1 \over \tilde\lambda}\sqrt{{M \over \lambda}}
\e^{-\tilde\lambda t }\sinh \tilde\lambda r}
\eqn\exii$$
and Wick rotating $t \longrightarrow i\tau$.
The resulting metric is given by
$$ds^2=dr^2+\tanh^2\tilde\lambda r\, d\tau^2\ .\eqn\eixx$$
The geometry approaches to Euclidean flat space-time when $r$ goes to
$\infty$, $ds^2\sim dr^2+d\tau^2$,
and the metric has a following form when $r\rightarrow 0$,
$$ds^2\sim dr^2+\tilde\lambda^2 r^2 d\tau^2 \ . \eqn\exx$$
Since the metric is not singular if and only if $\tau$ has a period of
${2\pi \over \tilde\lambda}$, the temperature $T$ is given by,
$$T={\tilde\lambda \over 2\pi}\ .\eqn\exxi$$
This equation \exxi\ tells that the extremal solutions, which correspond
to $\tilde\lambda^2=0$, have a vanishing temperature. This result should
compare with CGHS black holes, which have a common non-vanishing
temperature $\lambda$.
When $\tilde\lambda^2<0$, the temperature is imaginary and
the naked singularity appears.

If we fix $\tilde M=\tilde\lambda^2 M$ to be finite
and redefine the coordinates $x_\pm$ by,
$$x_+\rightarrow x_++{\alpha \tilde M \over \lambda \tilde\lambda^2}\ , \ \
x_-\rightarrow x_-+\beta+{1 \over \alpha\tilde\lambda^2}\ , \eqn\exxa$$
we obtain the following metric,
in the limit of $\tilde\lambda^2\rightarrow 0$,
$$\ephi=\erho={1 \over \alpha}x_++\alpha \lambda \tilde M x_-
+\beta\ .\eqn\exxb$$
The curvature is given by,
$$\eqalign{R&=8\erho\partial_+\partial_-\rho\cr
&={4\lambda \tilde M \over
{1 \over \alpha}x_++\alpha \lambda \tilde M x_-+\beta}\ ,}
\eqn\exxc$$
and we find that there appears a singularity when
${1 \over \alpha}x_++\alpha \lambda \tilde M x_-+\beta=0$.

The solutions \ex\ can also describe charged matter (chiral fermions)
collapsing into black holes. As an example, we can consider a charged
shock wave which is given by,
$$\eqalign{{i \over 4}\sum_{j=1}^N(\psi_j^*(x^+)\partial_+\psi_j(x^+)
-\partial_+\psi_j^*(x^+)\psi_j(x^+))&=a\delta(x^+-x^+_0)\ ,\cr
{g_A^2 \over 4}\sum_{j=1}^N\psi_j^*(x^+)\psi_j(x^+)&=b\delta(x^+-x^+_0) \ .}
\eqn\exxii$$
Then the solution is given by,
$$\eqalign{x^+<x^+_0\ , \ \ F_{+-}&=C\ ,\cr
\ephi&=\erho={M \over \lambda}-(\lambda^2-{2C^2 \over g_A^2})x^+x^-\ ,\cr
x^+>x^+_0\ , \ \ F_{+-}&=C+b\ ,\cr
\ephi&=\erho={M \over \lambda}+ax^+_0
+{2Bax^+_0 \over g_A^2\tilde\lambda'^2}\cr
&-\tilde\lambda'^2
(x^++{2Bx^+_0 \over g_A^2\tilde\lambda'^2})
(x^-+{a \over \tilde\lambda'^2})\ .}
\eqn\exxiii$$
Here $\tilde\lambda'$ and $B$ are defined by,
$$\tilde\lambda'^2\equiv\lambda^2-{2(C+b)^2 \over g_A^2}\ , \ \
B\equiv (2Cb+b^2)\ . \eqn\exxiiia$$
When $\tilde\lambda'=0$, the metric \exxii\ in case $x^+>x^+_0$ corresponds
to the metric in Equation \exxc .
Note that there appears a singularity even in case $\tilde\lambda'=0$.

Equation \exxiii\ tells that the event horizon when $x^+>x^+_0$ is given by
$$(x^++{2Bx^+_0 \over g_A^2\tilde\lambda'^2})
(x^-+{a \over \tilde\lambda'^2})=0\ .
\eqn\exxiv$$
Note that not only future horizon corresponding to
$x^-=-{2Bx^+_0 \over g_A^2\tilde\lambda'^2}$, but past horizon, where
$x^+=-{a \over \tilde\lambda'^2}$, is shifted by the
shock wave. In the original paper by 't Hooft\cref\v the shift of
only future horizon was discussed in four dimensional black holes.
The shift of the past horizon, which is observed in this paper,
would suggest that charged particles collapsing into \RN\ black hole
in four dimensions could shift the past horizon.
The meaning of the shift will be dicussed
in the forthcoming paper\pref\xiii

In the original paper by CGHS, they found that the Hawking radiation and
the back reaction of the metric can be described by adding correction
terms to the classcal equation of motion.
The correction terms come from the conformal anomaly.
Now, since we have chiral fermions, the terms which come from
the chiral anomaly should be also added. Then the total quantum action $S_q$
is given by,
$$\eqalign{S_q&=S_c+S_\rho+S_\chi\ ,\cr
S_\rho&=\pint\sqg\{-\half(\nabla Z)^2+\sqrt{{N \over 48}}ZR\}\ ,\cr
S_\chi&=\pint\{\half\sqg(\nabla Y)^2
+\sqrt{{N \over 2}}Y\epsilon^{\mu\nu}F_{\mu\nu}\}\ .}
\eqn\exxv$$
Here $S_c$ is a classical action in Eq.\ei.
By adding new degrees of freedom $Z$ and $Y$, we write the actions in local
forms.
By choosing the gauge fixing conditions \eii\ and \eiib ,
the classical equations of motion \ev\ - \eix\ are modified as follows,
$$\eqalign{0&=T_{++}=\ephi(4\partial_+\rho\partial_+\phi-2\partial_+^2\phi)
+\half\partial_+Z\partial_+Z-\sqrt{{N \over 12}}\partial_+\rho\partial_+Z
+{1 \over 2}\sqrt{{N \over 12}}\partial_+^2 Z\cr
& \ \ \ -\half\partial_+Y\partial_+Y
+{i \over 4}\sum_{j=1}^N(\psi_j^*\partial_+\psi_j-\partial_+\psi_j^*\psi_j)
+\half A_+\sum_{j=1}^N\psi_j^*\psi_j
\ , \cr
0&=T_{--}=\ephi(4\partial_-\rho\partial_-\phi-2\partial_-^2\phi)
+\half\partial_-Z\partial_-Z-\sqrt{{N \over 12}}\partial_-\rho\partial_-Z
+{1 \over 2}\sqrt{{N \over 12}}\partial_-^2 Z\cr
& \ \ \ -\half\partial_-Y\partial_-Y
\ , }
\eqn\exxvi$$
$$0=T_{+-}=\ephi(2\partial_+\partial_-\phi
-4\partial_+\phi\partial_-\phi-\lambda^2\eprho)
+{2 \over g_A^2}\e^{2\phi-2\rho}F_{+-}^2
-\sqrt{{N \over 48}}\partial_+\partial_-Z
\ , \eqn\exxvii$$
$$\eqalign{0&=-{8 \over g_A^2}\partial_+(\ephirho F_{+-})
-\sqrt{2N}\partial_+Y+\half\sum_{j=1}^N\psi_j^*\psi_j \ ,\cr
0&=-{8 \over g_A^2}\partial_-(\ephirho F_{+-})
-\sqrt{2N}\partial_-Y
\ . }\eqn\exxviii$$
The dilaton and matter equations of motion \eviii\ and \eix\ are not modified,
$$0=-4\partial_+\partial_-\phi+4\partial_+\phi\partial_-\phi
+2\partial_+\partial_-\rho+\lambda^2\eprho
-{2 \over g_A^2}\e^{4\phi-2\rho} F_{+-}^2
\ , $$
$$0=\partial_-\psi_j \ .$$
We also have $Z$ and $Y$ equations,
$$0=-2\partial_+\partial_-Z
+\sqrt{{N \over 3}}\partial_+\partial_-\rho \ ,\eqn\exxixa$$
$$0=2\partial_+\partial_-Y+\sqrt{2N}F_{+-} \ .\eqn\exxixb$$

The modified Maxwell equation \exxviii\ can be solved with respect
to $Y$ as follows
$$\sqrt{2N}Y=-{8 \over g_A^2}\ephirho F_{+-}
+\half\int^{x^+}dy^+\sum_{j=1}^N\psi_j^*(y^+)\psi_j(y^+)\ . \eqn\exxx$$
By substituting this equation \exxx\ into Equation \exxixb ,
we obtain,
$$0=\partial_+\partial_-(\ephirho F_{+-})-{g_A^2 N \over 8}F_{+-} \ .
\eqn\exxxi$$
This equation tells that electromagnetic fields become massive due to
the screening effects by fermion loops.
This screening effects make the field strength $F_{+-}$ decrease
in the asymptotic region and increase near the singularity.
Therefore we can expect that
$\tilde \lambda^2=\lambda^2-{2 \over g_A^2}\e^{4(\phi-\rho)}F_{+-}^2$
becomes $x^\pm$ dependent and could be negative in the horizon.
If the metric is effectively given by Equations \eii\ and
\exi\ even in quantum theory, the singularity could disappear due to this
screening effect.

Equation \exxixa\ can be also solved with respect to $Z$ as follows,
$$Z=\sqrt{{N \over 12}}\rho+r^+(x^+)+r^-(x^-) \ .\eqn\exxxii$$
By using this equation, Equation \exxvi\ can be written as follows,
$$\eqalign{0&=\ephi(4\partial_+\rho\partial_+\phi-2\partial_+^2\phi)
-{N \over 24}(\partial_+\rho\partial_+\rho-\partial_+^2 \rho)+t^+(x^+)\cr
& \ \ \ -{1 \over 4N}\{\partial_+(-{8 \over g_A^2}\ephirho F_{+-}
+\half\int^{x^+}dy^+\sum_{j=1}^N\psi_j^*(y^+)\psi_j(y^+))\}^2\cr
& \ \ \
+{i \over 4}\sum_{j=1}^N(\psi_j^*\partial_+\psi_j-\partial_+\psi_j^*\psi_j)
+\half A_+\sum_{j=1}^N\psi_j^*\psi_j\ ,\cr
0&=\ephi(\partial_-\rho\partial_-\phi-2\partial_-^2\phi)
-{N \over 24}(\partial_-\rho\partial_-\rho-\partial_-^2 \rho)+t^-(x^-)\cr
& \ \ \ -{1 \over 4N}\{\partial_-(-{8 \over g_A^2}\ephirho F_{+-})\}^2
\ . }
\eqn\exxxiii$$
Here $t^\pm(x^\pm)$ is defined by
$$t^\pm(x^\pm)\equiv \half(\partial_\pm r^\pm)^2
+{1 \over 4}\sqrt{{N \over 3}}\partial_\pm^2r^\pm \ . \eqn\exxxiv$$
Furthermore we obtain the following equations by using Equations
\exxvii , \eviii\ and \exxxii ,
$$(1-{N \over 48}\epphi)\partial_+\partial_-\rho
=\partial_+\partial_-\phi \ , \eqn\exxxiva$$
$$2(1-{N \over 24}\epphi)\partial_+\partial_-\phi
=(1-{N \over 48}\epphi)\{4\partial_+\phi\partial_-\phi
+\eprho(\lambda^2-{2 \over g_A^2}\e^{4(\phi-\rho)}F_{+-}^2)\}\ .
\eqn\exxxivb$$
The classical solution describing the dilaton vacuum \exa\ satisfies Equations
\exxxi , \exxxiva\ and \exxxivb .
If we require that the solution also satisfies \exxxiii ,
we find that $t^\pm(x^\pm)$ in Equation \exxxiii\ should be given by,
$$t^\pm(x^\pm)=-{N \over 96 (x^\pm)^2} \ .\eqn\exxxv$$

By following the paper by J.G. Russo, L. Susskind and L. Thorlacius\cref\vi
we consider the solution describing a charged shock wave collapsing into
the dilaton vacuum \exa .
On the matter trajectory $x^+=x^+_0$, we consider the differential equations
with respect to $\Sigma\equiv\partial_+\phi$ with the following
functional coefficients
$$\ephi=\erho=-\lambda^2x^+_0x^-\ ,\ \ F_{+-}=C\ .\eqn\exxxvi$$
Note that the value of $F_{+-}$ jumps by the charged shock wave.
Then the equation \exxxivb\ becomes on the trajectory $x^+=x^+_0+0$
$$2(1+{N \over 24\lambda^2x^+_0x^-})\partial_-\Sigma
=(1+{N \over 48\lambda^2x^+_0x^-})(-{2 \over x^-}\Sigma
-{\tilde\lambda^2 \over \lambda^2x^+_0x^-})
\eqn\exxxvii$$
Here $\tilde\lambda$ is defined by Equation \exiii .
The solution of this equation is given by,
$$\Sigma=-{\tilde\lambda^2 \over 2\lambda^2}\{{1 \over x^+_0}
-{{M \over \lambda x^+_0} \over \sqrt{-\lambda^2x^+_0x^-(-\lambda^2x^+_0x^-
-{N \over 24})}}\} \ . \eqn\exxxviii$$
The integration constant is fixed by the condition that $\Sigma$ approachs
$\partial_+\phi$ of the classical black hole solution \exxiii\
as $x^-\rightarrow -\infty$.
Analogously to CGHS case\cref\vi there appears a singularity at
$x_-=-{N \over 24\lambda^2x^+_0}$ except case $\tilde\lambda=0$, \ie ,
the case that the resulting black hole is extremal.
The curve of singularity is given by $1={N \over 24}\epphi$. In order to
investigate the behavior of the curve, we consider the curves $x^-=f(x^+)$
where $\phi$ is a constant. Since
$$0={d\phi \over dx^+}\mid_{x^-=f(x^+)}=\partial_+\phi+f'\partial_-\phi\ ,$$
$f'$ is given on the trajectory $x^+=x^+_0$,
$$f'=-{\Sigma \over \partial_-\phi}
={\tilde\lambda^2 x^- \over \lambda^2 }\{{1 \over x^+_0}
-{{M \over \lambda} \over \sqrt{-\lambda^2x^+_0x^-(-\lambda^2x^+_0x^-
-{N \over 24})}}\} \ , \eqn\exxxviiib$$
which is negative near the singularity. Note that $f'$ is proportional to
$\tilde\lambda^2$. Now we assume the dominant part of $f'$ would be
proportional to $\tilde\lambda^2$ even in the region $x^+>x^+_0$.
Since $\tilde\lambda^2$ would increase exponentially due to the screening
effect as $x^+$ increases, the sign of $f'$ would
turn to be negative and after that, its absolute value $|f'|$ would also
increase exponentially and $x^-$-coordinate of the singularity would go to
infinity. Therefore the decaying black hole could evolve into an object
without singularity.
More definite argument will be given elsewhere\pref\xiii

Although we have a classical solutions which describe extremal solution
with the vanishing temperature, we cannot find if the solution can be
a remnant of black holes. In order to obtain more definite answer,
we need to solve the following equations:
$$\eqalign{0&=\partial_+\partial_-(\ephirho F_{+-})
-{g_A^2 N \over 8}F_{+-}\ , \cr
0&=\ephi(4\partial_+\rho\partial_+\phi-2\partial_+^2\phi)
-{N \over 24}(\partial_+\rho\partial_+\rho-\partial_+^2 \rho)+t^+(x^+)\cr
& \ \ \ -{1 \over 4N}\{\partial_+(-{8 \over g_A^2}\ephirho F_{+-})\}^2\ , \cr
0&=\ephi(4\partial_-\rho\partial_-\phi-2\partial_-^2\phi)
-{N \over 24}(\partial_-\rho\partial_-\rho-\partial_-^2 \rho)+t^-(x^-)\cr
& \ \ \ -{1 \over 4N}\{\partial_-(-{8 \over g_A^2}\ephirho F_{+-})\}^2\ ,\cr
0&=(1-{N \over 48}\epphi)\partial_+\partial_-\rho
-\partial_+\partial_-\phi \ , \cr
0&=2(1-{N \over 24}\epphi)\partial_+\partial_-\phi
-(1-{N \over 48}\epphi)\{4\partial_+\phi\partial_-\phi
+\eprho(\lambda^2-{2 \over g_A^2}\e^{4(\phi-\rho)}F_{+-}^2)\}\ , }
\eqn\exxxix$$
with the boundary conditions \exxxvi\ at $x^+=x^+_0$.
Equations \exxxix\ are obtained by setting $\psi_j=0$ in Equations
\exxxi , \exxxiii, \exxxiva\ and \exxxivb .

The $U(1)$ gauge symmetry can break down
spontaneously by adding the Higgs fields action $S_H$ to the original
action \ei ,
$$S_H=\pint\sqg g^{\mu\nu}(\partial_\mu\varphi-A_\mu)
(\partial_\nu\varphi-A_\nu)\ .\eqn\exxxx$$
In the resulting theory, any charged black hole does not appear
since the gauge fields $A_\mu$ become massive.
If we fix the $U(1)$ gauge symmetry by the gauge condition
$$\varphi=0\ , \eqn\exxxxi$$
the general solutions in the conformal gauge \eii\ are given by,
$$\eqalign{
A_+&=\int da af(a)\e^{{i \over 2\sqrt 2}(ag_Ax^++{g_A \over a^*}x^-)}\ , \cr
A_-&=-\int da {1 \over a}f(a)
\e^{{i \over 2\sqrt 2}(ag_Ax^++{g_A \over a^*}x^-)}\ , \cr
\ephi&=\erho={M \over \lambda}-\lambda^2x^+x^-\cr
&-{1 \over 2g_A^2}\int da \int db {abf(a)f(b) \over (a+b)^2}
\e^{{i \over 2\sqrt 2}\{(a+b)g_Ax^++g_A({1 \over a^*}+{1 \over b^*})x^-)}\ .}
\eqn\exxxxii$$
Here $a$ can be imaginary in general but we choose a function $f(a)$
and the region of integration with respect to $a$ in order for $A_\pm$
to be real.
Coleman, Preskill and Wilczek (CPW) have considered quantum hairs
by breaking $U(1)$ gauge symmety by charge $n$ particles\pref\x
In their scenario, there remains $Z_n$ gauge symmetry and the Hilbert space
decomposes into $n$ sector. They expected that the
decomposition affects the temperature of the black hole.
The $Z_n$ charge gives a quantum hair and this hair
can be detected through topological quantum effect.
Since the topological effects in two dimensions differ from those
in four dimensions, we cannot simply apply their argument to the model
proposed
here.
We expect, however, that
some effects coming from the decomposition might be detected even in two
dimensions.


We acknowledge K. Odaka and A. Sugamoto for discussions.
We are also indebted to S. Odake and T. Tada for useful informations.
The research of I.O. is supported by the Japan Society for the Promotion
of Science.

\refout

\bye